# Resequencing: A Method for Conforming to Conventions for Sharing Credits Among Multiple Authors

Ash Mohammad Abbas
Department of Computer Engineering
Aligarh Muslim University
Aligarh - 202002, India.

*Abstract*—Devising an appropriate scheme that assigns the weights to share credits among multiple authors of a paper is a challenging task. This challenge comes from the fact that different types of conventions might be followed among different research discipline or research groups. In this paper, we discuss that for the purpose of evaluating the quality of research produced by authors, one can resequence either authors or weights and can apply a weight assignment policy which the evaluator deems fit for the particular research discipline or research group.

## I. Introduction

Sometimes, one is required to evaluate the quality of research produced by an author or a set of authors. In case of multi-authored papers, one should consider the contribution of each author and for that purpose one should assign weights to each author of the paper. However, devising an appropriate scheme that assigns the weights for sharing credits among multiple authors is a challenging task and the challenge comes due to the fact that different types of conventions are followed among different research disciplines and/or research groups. For example, in some of the disciplines, the first author and the last author of a paper are considered to be the lead authors and rest of the authors are assumed to contribute less as compared to these authors. In some of the disciplines (or groups), authors are considered to contribute in the order in which their names appear in the list of authors of the paper. In some disciplines, it might be the corresponding author who is assumed to contribute to the major part of the paper, irrespective of the position where his/her name appears in the list of authors of the paper.

There is no universally agreed weight assignment scheme for sharing credits among multiple authors. One scheme of assigning weights to authors can be that each author is treated to contribute equally to the research paper. This is called *fractional* or *equal weight assignment scheme*. It has been pointed out in [3] that the trend of equally sharing the credits is increasing. As the weight assignment in a fractional scheme does not depend upon the position of authors, therefore, there is no issue related to the relative contributions of authors. However, practically the contributions of authors are not equal [4] and there are no means to measure the contributions of individual authors. Though, some journals require that authors should themselves declare who carried out the research, who has written the paper, etc, in order to identify and reflect the contributions of individual authors.

Further, in research team some times there is mix of authors, some of them are quite efficient and contributed to the paper in large part, and some of them contributed for portions of the work, however, their contributions might not be so less that one can leave them only acknowledging their help. It may also happen that in some other papers the authors who contributed less in one paper might have contributed relatively a large portion of the work in a collaboration with some other authors. An equal weight assignment policy may not be applied in such a case. Therefore, it seems logical that if the contribution of authors are not more or less equal, there should be a scheme that takes into account the contributions of individual authors. As mentioned above, there is no scheme which can be applied in all research areas or research groups due to different types of conventions followed among different research disciplines or research groups. The conventions may depend on social economic, political, academic factors.

In our previous works, we discussed weighted indices to incorporate multiple authorship [1], and generalized linear weights to share credits among multiple authors [2]. In this paper, we discuss that in case of an unequal weight assignment to multiple authors one can apply a weight assignment scheme which he/she considers an appropriate scheme for the research discipline or research group. To make the scheme fit to the order of authors, one can use resequencing of either authors or weights.

The rest of this paper is organized as follows. In section II, we describe reordering of authors and weights. In section III, we discuss using examples how one can reorder the authors and weights in different scenarios. In section IV, we discuss the usability of resequencing. Finally, in the last section we conclude the paper.

## II. Resequencing

In this section, we describe what we mean by resequencing. The resequencing can be of two types: (i) *author resequencing*, and (ii) *weight resequencing*. We describe both of them as follows.

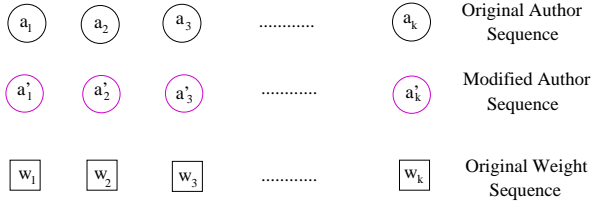

Fig. 1. The authors are resequenced, however, the sequence of weights from the first authors to the last author remains the same.

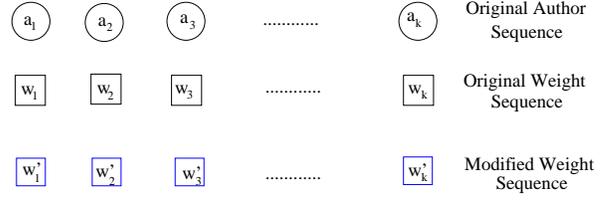

Fig. 2. The weights are resequenced, however, the sequence of authors from the first authors to the last author remains the same.

## A. Author Resequencing

Let there be $k$ number of authors of a paper in the following order $<a_1, a_2, ..., a_k>$, and let there be a weight assignment scheme, $\mathcal{S}$, that assigns the following weights to the set of authors $<w_1, w_2, ..., w_k>$ in that order. By resequencing the authors, we mean that the authors are rearranged according to the conventions followed in the research discipline or research group, however, the order of weights remains the same as that provided by the weight assignment policy. Let the new order of authors be $<a'_1, a'_2, ..., a'_k>$. It means that the weight $w_i$ is assigned to author $a'_i$ for $i = 1, ..., k$. Figure 1 shows author resequencing.

## B. Weight Resequencing

In weight resequencing, the order of author remains the same as in the list of authors. However, the weights provided by the weight assignment policy are reordered using the conventions followed by a research discipline or a research group. Let the sequence of authors is $<a_1, a_2, ..., a_k>$ and the sequence of weights provided by the weight assignment scheme, $\mathcal{S}$, be $<w_1, w_2, ..., w_k>$. Let the set of weights after resequencing be $<w'_1, w'_2, ..., w'_k>$. It means that the weight $w'_i$ is assigned to the author $a_i$. Figure 2 shows the resequencing of weights.

## III. EXAMPLES OF SCENARIOS FOR RESEQUENCING

We now discuss examples of different types of scenarios where resequencing may be required to share credits among multiple authors. These scenarios are based on the conventions used in a research discipline or a research group.

Let the convention followed by a research discipline or a research group, say A, be as follows.

*Convention 1:* The last author is the lead author and should be assigned the largest weight amongst all authors of the paper.

*Solution 1:* Since the last author is the lead author, therefore, it can be brought at the first place, and the rest of the authors are shifted to the right by one place. In other words, the sequence of authors after reordering is as follows.

$$<a_k, a_1, a_2, ..., a_1>.$$

This is resequencing the authors, the weights remain the same as provided by the a weight assignment scheme, $\mathcal{S}$. Note that the new sequence of authors is shifted to the right in a circular fashion.

A similar solution can be obtained using resequencing of weights. If the weights provided by the weight assignment scheme, $\mathcal{S}$, are in descending order, then the weight of the first author can be put at the last place and weights of authors are shifted to left by one place. The new set of weights is as follows.

$$<w_2, w_3, ..., w_k, w_1>.$$

Again, we remind that weight reseuencing in this case is shifting the weights to the left in a circular fashion.

Let us consider another convention followed by a research discipline or a research group, say $B$, as follows.

*Convention 2:* The last author of the paper is the group leader, and his contribution to the paper is less than the first author, however, greater than any other author.

*Solution 2:* Let us first consider author resequencing. As mentioned earlier, the weights generated by a weight assignment scheme, $\mathcal{S}$, are in descending order. In the modified sequence of authors, the first author should be at the first position which corresponds to the largest weight, the last author should be at the position corresponding to the second largest weight and it is actually the second position, the rest of the authors are placed starting from third position to the last position. In other words, the new sequence of authors is as follows.

$$<a_1, a_k, a_2, a_3, ..., a_{k-1}>$$

On the other hand, consider the weight resequencing. The first author, $a_1$, should be assigned the largest weight, $w_1$, the last author, $a_k$, should be assigned the second largest weight, $w_2$, and the rest of the authors should be assigned the remaining weights in decreasing order. In other words, the new sequence of weights is as follows.

$$<w_1, w_3, w_4, ..., w_k, w_2>.$$

Let us consider another convention, which has a notion of the corresponding author, and is followed by a research discipline or a research group, say $C$, as follows.

*Convention 3:* The corresponding author of the paper is assigned the largest weight, then the first author, and then the last author; the rest of the authors have their contribution in the order of their names appearing in the author list.

*Solution 3:* Let $j$th author be the corresponding author, then reseuencing the authors gives the following new sequence of authors.

$$<a_j, a_1, a_k, a_2, ..., a_{j-1}, a_{j+1}, ..., a_{k-1}>.$$

Similarly, one can work out the sequence of weights if one wishes to use weight resequencing as follows. The largest weight, $w_1$, is assigned to the corresponding author, $a_j$, the second largest weight, $w_2$, is assigned to the first author, $a_1$, and the third largest weight, $a_3$, is assigned to the last author, $a_k$. In other words, the set of weights after resequencing them is as follows.

$$< w_2, w_4, w_5, ..., w_{j-1}, w_1, w_{j+1}, ..., w_{k-1}, w_3 >.$$

Similarly, there can be other scenarios, and one can think of how one should resequence either the list of authors or the set of weights which is appropriate for a particular research discipline or a research group.

Note that there can be situations where we will not recommend the use of any of the resequencing discussed in this paper. An example of such a situation can be as follows. The names of authors are listed in alphabetical order, and there is no explicit declaration about the specific contributions of authors. In the absence of such a declaration, one should consider an equal weight assignment scheme. Actually, the resequencing of either authors or weights is for the situations where the contributions of authors are unequal and a weight assignment scheme which generates unequal weights is used. If authors were in alphabetical order, however, there is an explicit declaration about the contributions of authors; and from the declaration, one is able to decide an unequal extent of contribution by the authors of the paper, then one can consider to use an unequal weight assignment scheme, in that case, one can consider to apply the resequencing of either weights or authors.

## IV. Applications of Resequencing

As mentioned earlier, one can consider to use the resequencing of either authors or weights provided that one has decided to use an unequal weight assignment scheme. The modified sequence of weights or authors can be used to compute an index. Let there be an index $I$ that uses the number of citations of papers authored by a researcher in a manner that is specific to the index. The citations are considered to be the credits for authors of the papers. In case when there are more than one authors of a paper, the credits (or the number of citations) should be divided among the authors using an appropriate weight assignment scheme. Let $c_i$ be the number of citations of $i$th paper of an author and $w_i$ be the weight assigned to the author for his/her $i$th paper, then the number of weighted citations of the given author for his/her $i$th paper is as follows.

$$c'_i = c_i w_i.$$

The weighted number of citations for $i$th paper of the given author, $c'_i$, should be used for computing a given index, say $I$, following the procedure of the index $I$.

This can be done for the purpose of evaluating the quality of research produced by an author. The type of resequencing used depends on the evaluator. Further, we would like to mention that the resequencing of either authors or weights can be incorporated in an indexing database. An evaluator (or the end-user) can be allowed to resequence either the authors or the weights and then should be able to compute an appropriate index. For that purpose, there can be set of default weights provided by the indexing database, and the user, if he/she wishes, can be allowed to resequence the set weights.

## V. Conclusion

Finding an appropriate weight assignment scheme to share credits among multiple authors of a paper is a challenging task due to the fact that the conventions followed among different research disciplines or research groups. In this paper, we described that for an unequal weight assignment scheme that generates weights in a descending order, one can use resequencing of either authors or weights so as to conform to the conventions followed by a research discipline or a research group. This can be done for the purpose of evaluating the quality of research produced by an author, however, the type of resequencing used depends on the evaluator, and the evaluator can be asked to select a resequencing method by an indexing database.